\documentclass[11pt]{article}
\usepackage{iap2000,epsfig}

\def\Journal#1#2#3#4{{#1} {\bf #2}, #3 (#4)}

\def\AA{{\em A\&A,}}
\def\AAS{{\em A\&AS,}}

\def\PRE{{\em Phys. Rev.} E}

\def\AASS{{\em A\&A Suppl. Ser.,}}
\def\APJ{{\em ApJ,}}
\def\APJS{{\em ApJS,}}
\def\APJL{{\em ApJL,}}
\def\ARA&A{{\em Ann. Rev. A\&A,}}
\def\AJ{{\em AJ,}}
\def\MNRAS{{\em MNRAS,}}
\def\The Messenger{{\em The Messenger,}}

\def\be{\begin{equation}}
\def\ee{\end{equation}}
\def\bea{\begin{eqnarray}}
\def\eea{\end{eqnarray}}

\def\deg{\hbox{$^\circ$}}
\def\sun{\hbox{$\odot$~}}
\def\arcmin{\hbox{$^\prime$}}
\def\sq{\hbox{\rlap{$\sqcap$}$\sqcup$}}
\def\lefevre{Le\thinspace F\`evre}
\def\etal{{\it et al.} }
\def\hawaii{Hawai$'$i}
\def\arcsec{\ifmmode^{\prime\prime}\;\else$^{\prime\prime}\;$\fi}

\begin{document}
\vspace*{4cm}
\title{CLUSTER SURVEYS}

\author{Isabella M. GIOIA}

\address{Istituto di Radioastronomia del CNR, Bologna, 
Italy \\
Institute for Astronomy, Honolulu, \hawaii, USA}

\maketitle\abstracts{
I review some of the current efforts to create unbiased samples
of galaxy clusters. Readers are referred elsewhere for general wide 
area sky surveys and redshift surveys, and for Sunyaev-Zeldovich, radio, 
infrared and submm surveys, some of which were not designed to search 
primarily for clusters. My focus will be on optical and X-ray 
samples and on high redshift clusters. 
}

\section{Importance of Galaxy Clusters}
Clusters of galaxies are the largest virialized, bound systems known.
Their study yields a wealth of information on structure in the 
Universe over a large range of length scales and on the evolution  
of that  structure over a long time span. Individual clusters are 
the ideal laboratory for multiwavelength
studies of many physical processes involving various constituents of
the Universe (galaxies, intracluster hot gas, magnetic fields and 
relativistic particles) and the effect of the environment 
on their evolution. On larger scales, samples of clusters are very 
useful tools to understand
the formation and evolution of large-scale structure through the study
of the average statistical properties of clusters as a function of
redshift. Since clusters are relatively luminous at many wavelengths, they 
may be more easily traced to very high redshift (even though there are not
many clusters known beyond $z\geq1$) and thus they represent a very
important observational tool for cosmologists. In particular, they can 
be used to constrain cosmological parameters.

All this explains the past and present efforts made in cluster surveys
to build very large cluster samples and to push the detection of
clusters to higher and higher redshift. In this respect the search for 
clusters has greatly benefitted from new technologies and observational 
capabilities. Faint spectroscopic data from Keck and VLT, deep optical 
and near-infrared imaging in space (HST) and from the ground, observations
of the intracluster medium with ROSAT and ASCA, enhanced imaging and 
spectroscopic capabilities of Chandra and XMM-Newton, have considerably 
added to our knowledge of these objects.

Cluster searches are conducted in different
waveband domains. The traditional searching waveband for clusters
has been the optical wavelength, but in the last twenty years X-ray searches
have played an increasingly important role. More recently 
radio, infrared and submm searches 
have also been succesfully used to detect bright clusters.
All cluster finding methods, in any waveband domain, suffer 
from selection effects and biases. What is important is to be aware 
of the selection effects and to be able to correct for the biases.
It is also desirable to have catalogs extracted from as many different
wavelength domains as possible, since each sample will contain
a somewhat different cluster population with different selection 
effects. Here is an (incomplete) list of all-sky or extra-galactic 
surveys carried out at different wavelengths. Most of these surveys 
appeared in the literature in the past decade. Cluster studies have 
greatly benefitted from these surveys
even if several of them were not specifically designed to search 
for such systems.

POSS and SERC (and related digitized surveys)

ROSAT \cite{vog99}

NVSS \cite{con98}

FIRST \cite{bec95}

WENSS \cite{ren97}

IRAS \cite{rr90}

IRAS PSCz Redshift Survey  \cite{sau98}

CFA redshift survey \cite{huc82}$^{,}$ \cite{del86}$^{,}$ \cite{huc99}

2dFGRS \cite{col99}

6dF \cite{col00}

LCRS \cite{she96}

NBG groups of galaxies \cite{tul87}

VIRMOS \cite{lfev00}

ESP \cite{vet97} 

SUMSS \cite{boc99}

2MASS \cite{jar00}

DENIS \cite{mam98}

SDSS \cite{kna99}$^{,}$ \cite{ma99} 

Readers are referred elsewhere for details on each
survey. Here I will mention only the Sloan Digital Sky Survey 
(SDSS), which is an ambitious project involving a joint 
venture of several astronomical institutes.

\subsection{The Sloan Digital Sky Survey (SDSS)}
The Sloan Digital Sky Survey\cite{kna99} will eventually produce a 
map of 10,000\ \sq\deg\ of the northern sky and 225\ \sq\deg\ of the
southern  sky. The SDSS is both a photometric (five-bands: u\arcmin, 
g\arcmin, r\arcmin, i\arcmin, z\arcmin) and a spectroscopic survey 
which will permit the selection of galaxy clusters using well-defined, 
automated algorithms. Cluster investigations will greatly benefit from the 
SDSS data, either alone or in combination with other wavelength 
datasets, from measurements of cluster parameters to correlations 
between cluster properties and their evolution. 
The very large volume and objective identification methods of the SDSS 
will result in a uniform  database of objects including an enormous 
sample of clusters. This sample can be used to characterize the 
cluster population statistically with great accuracy. It will also
allow to test the cosmological models through the measure of
galaxy clustering and to characterize the large-scale structure at the 
present epoch ($z<0.2$).

\section{Cluster Surveys and Catalogs in Optical}
In the years following the pioneering work of Abell \cite{abe58}, 
only a few surveys were produced to search systematically 
for clusters of  galaxies. The early generation 
systematic surveys  \cite{abe58}$^{,}$ \cite{abe89}$^{,}$  
\cite{zwi68} were done by visual inspection of plates and thus
with no quantifiable selection functions. The Abell catalog was 
for many years the only cluster catalog available
and thus heavily used for cluster studies,  including 
discovery of superclusters, and large-scale structure
investigations up to $z\leq0.2$. At higher redshift ($z<0.5$) 
surveys based on prime focus deep plates \cite{gho86}$^{,}$ \cite{cou91}
have provided information on galaxy evolution, on galaxy
clustering and space densities as a function of richness,
on the Butcher-Oemler effect, etc. But the real breakthrough was
the emergence of optical cluster catalogs with completely automated  
selection and quantifiable selection criteria (see among others
\cite{she85}$^{,}$ \cite{lum92}$^{,}$ \cite{dal92}).
The automated  surveys increased the efficiency of cluster searches and 
allowed more accurate determinations of cluster space density 
measurements. At even higher redshifts ($z\geq0.5$)
it becomes extremely difficult to detect enhancements in the galaxy 
surface density against the overwhelming field galaxy population
and optical surveys are known to suffer from effects of superpositions 
of unvirialized systems (see among others  \cite{fre90} and \cite{vh97}).
Essentially all optical surveys still use richness, either in 
two dimensions or three dimensions, as the selection criterion, but richness 
correlates poorly with mass because of the projection effects.
Fully automated and objectively selected catalogs of clusters 
with quantifiable selection criteria (different from the traditional 
$\delta\rho/\rho$) are thus necessary for these catalogs to 
be statistically useful.

I will discuss now current efforts that should have a significant 
impact on optical/NIR cluster studies. 

\subsection{The Palomar Distant Cluster Survey (PDCS)}
The number of known distant clusters has greatly increased  since 
the publication in 1996 of the PDCS catalog by Postman and 
collaborators \cite{pos96}.
The PDCS is among the first fully automated objectively selected 
catalog of clusters based on CCD imaging data. The investigators  
use a matched filter algorithm, in both positional and 
photometric data,
to search for overdensities in the galaxy distribution.
The 79 clusters and candidates in the original survey 
paper \cite{pos96}, of which 16 were already known from the 
catalog of Gunn, Hoessel and Oemler \cite{gho86}, cover the redshift 
range $0.2<z<1.2$. One of the main results of the survey is that
the PDCS cluster space density is a factor of about five above that 
seen in the local universe as measured from the Abell catalog.
Recently this result was confirmed by a spectroscopic survey of 
the PDCS clusters in the redshift range $0.1<z<0.35$ \cite{hol99}. 
Actually it was found 
that a fraction ($\sim$1/3) of the PDCS clusters have lower velocity 
dispersions (200 km s$^{-1}$, typical of groups) but richness 
appropriate for clusters. Excluding those low velocity dispersion 
clusters the PDCS cluster density estimate is of $\sim$ 3 times that of 
the Abell catalog for equivalent mass clusters of galaxies, in agreement
with the APM \cite{mad90}$^{,}$ \cite{dal97} and the Edinburgh-Durham 
cluster catalog \cite{bra00}, thus raising the possibility that the 
Abell catalog is incomplete. These results are different from the
results of the X-ray surveys which find incompleteness at a given
luminosity to be only 30\% \cite{eb98}$^{,}$ \cite{eb00a}$^{,}$
\cite{dg99}.

Oke and collaborators \cite{oke98} performed extensive spectroscopic 
observations with Keck to get spectra for 9 distant PDCS clusters ($z>0.6$). 
They measured over $\sim100$ redshifts per cluster for galaxies 
with R$\leq23.3$. A contamination of 30\% was found in the 
spectroscopic survey which led the authors to conclude that optical 
detection of clusters remains a successful and important method for 
identifying such systems out to $z\sim1$  (even if it suffers from 
projection effects) and that it will provide an important complement 
to cluster searches at other wavelengths. 
Recent findings from the Keck spectroscopic survey of the
9 PDCS clusters include the discovery of a group-group merger at 
z=0.84 \cite{lub98} and the optical detection of
a supercluster at $z\sim0.91$ \cite{lub00}.

\subsection{The ESO Imaging Survey (EIS) Clusters}
The matched-filter algorithm of Postman and collaborators
was applied to the EIS \cite{rdc97}$^{,}$ \cite{dc00} database, 
in its original version (see \cite{ols99a}$^{,}$ \cite{ols99b}$^{,}$ 
\cite{sco99}) and in a modified version (see \cite{lob00}).
The EIS is a public imaging survey, carried out by ESO and its 
community, to provide target lists for the Very Large Telescope. 
Among the goals of the EIS was the construction of a 
catalog of distant cluster candidates over a wide range in redshift 
from which targets could be drawn for subsequent follow-up studies. 
Different groups \cite{lob00}$^{,}$ \cite{ram00}$^{,}$ \cite{ols00}
are involved in the spectroscopic follow-up to confirm the cluster 
nature through spectroscopy. Preliminary results \cite{sco99}
show overdensities of red objects detected in over 300 group and cluster
candidates in 17\ \sq\deg. The estimated redshifts are in the range 
$0.2<z<1.3$ with a median redshift of $z\sim0.5$. This is the largest
sample of medium-distant clusters available in the southern hemisphere.

\subsection{The Low-Surface-Brightness Fluctuations in the Background Sky}
An interesting approach to find distant clusters is the one adopted
by Dalcanton \cite{jdc96} and by Zaritsky and collaborators
\cite{zar97}. These investigators use an old idea by Shectman
\cite{she73}$^{,}$ \cite{she74}, that the optical background 
light can be used to obtain information about the underlying 
distribution of galaxies.
They detect the light from the unresolved  galaxies in a cluster 
which manifest themselves as low-surface-brightness  fluctuations.
Extremely accurate flattening of the sky is required, but can be 
achieved with telescopes operating in drift-scan mode (like the great 
circle camera on the Las Campanas 1m telescope). The advantage of this
technique is that many square degrees of sky can be surveyed  in only 
a few nights using relatively small telescopes since dramatically 
shorter exposure time than for traditional surveys is required.
Of course deeper follow-up imaging is necessary for confirmation of 
the candidates plus spectroscopy for redshift determination.
An initial set of 10 cluster candidates, out of a preliminary list 
of 52 objects from a drift-scan survey (the Palomar Transit Grism Survey
\cite{jdc96}) of only 17.5\ \sq\deg\ of the northern sky, were 
confirmed as bona-fide clusters through imaging and spectroscopy.
The cluster  redshifts range from 0.4 to 1.06.
From a larger southern survey ($\sim130$\ \sq\deg\ of the sky
performed at Las Campanas Observatory) a catalog of over 1000 candidate 
clusters and groups has been produced. The latest results of the 
LCO survey are described by Dennis Zaritsky in these proceedings.

\subsection{Toronto Red-Sequence Cluster Survey (RCS)}
The Red-Sequence Cluster Survey (RCS) \cite{gy00}
is motivated by the observation that all rich clusters have
a population of early type galaxies which follow a strict 
color-magnitude relation. While the properties of the overall 
cluster galaxy population do evolve with redshift
(i.e. the blue fraction is generally higher at higher redshift), 
in all well-formed clusters so far observed there is a 
red sequence population not generally found in the field when 
looking over a large redshift range. Thus the searching 
algorithm is constructed to  exploit the red sequence of 
early-type galaxies as a direct cluster indicator.
The method has been tested using both 
real redshift survey data and thorough simulations. 
The RCS is designed to provide a large sample of optically selected 
clusters between $0.1<z<1.4$ by imaging 100\ \sq\deg\ in two filters 
(R and z\arcmin) to a depth sufficient to find galaxy clusters at 
$z\sim$1.4. The RCS survey  observations started in May of 1999, 
but initial analysis of  about 6\ \sq\deg\ of data has already 
allowed the discovery of numerous cluster candidates that can be
viewed at http://www.astro.utoronto.ca/$\sim$gladders/RCS/.

\subsection{The Canadian Network for Observational Cosmology (CNOC)}
A particular mention goes to the CNOC projects.
The CNOC survey \cite{yee96}$^{,}$ \cite{car96} consists of 
spectrophotometric observations of 15 intermediate
redshift and high luminosity EMSS clusters  
($0.1<z<0.6$; $L_{X}>4\times10^{44}$ erg s$^{-1}$) and, as such, it is 
not an effort aimed at finding new clusters of 
galaxies. The overall goal was to measure the total mass and 
luminosity in the central virialized region of the clusters to 
establish the value of $\Omega$ and measure any biases among cluster 
galaxies, cluster mass, and field galaxies. 
The survey developed into two separate projects: the CNOC1, 
which is a galaxy cluster redshift survey containing
over 2600 velocities used to measure Omega through the cluster dynamics; 
and the CNOC2 \cite{yee98} which is a field galaxy redshift survey 
containing about 6000 velocities to measure the evolution of 
galaxy clustering and galaxy populations. Both surveys are  used 
by the scientific community at large for cosmological studies.

\subsection{Radio and Near-Infrared Selection}
It was realized several years ago that luminous
radio galaxies at redshift $z\sim0.5$ inhabit rich 
clusters \cite{hl91}. There is now growing evidence that 
at higher redshifts ($z\geq3$) powerful radio sources are located 
at the center of forming clusters \cite{pen99}. Observational 
indications manifest as detection of luminous extended X-ray 
emission around the radio galaxies \cite{dic94}$^{,}$ \cite{cra96},
overdensities selected by infrared color \cite{dic97}, searches
around powerful Ultra-Steep Spectrum high-redshift radio galaxies 
\cite{cha96a}$^{,}$ \cite{cha96b}$^{,}$ \cite{pen97}, 
spectroscopic studies \cite{dic97}, detection of companion
galaxies with narrow-band imaging \cite{mcc95}, large Faraday 
depolarization and rotation measures of the radio sources 
\cite{best98} and so on.  Two examples are: MRC 0316-257 at 
z=3.12 \cite{lfev96}, 1138-262 at z=2.2 \cite{pen97}$^{,}$ \cite{kur00}. 
Other searches have shown overdensities of distant objects in fields
around weak steep spectrum radio galaxies (e.g. 53W002 at z=2.390 
\cite{pas96}$^{,}$ \cite{kee99}). From these observational results
it is clear that searches around high redshift radio galaxies 
are a good alternative to optical searches for finding very distant 
clusters of galaxies. Another interesting alternative to locate 
high redshift structures  are searches in fields of multiple lens 
quasars, as demonstrated by the triple quasar MG2012$+$112 at z=3.26, 
where a distant cluster at  z$\sim$1 has recently been 
discovered \cite{sou00}. However, it must be noted that these
techniques do not provide statistically complete samples of distant 
clusters.

In more recent years the near infrared has emerged as a new efficient band
to find distant clusters. Stanford \etal (1997) \cite{sta97} have shown 
that optical-IR colors (very red J-K colors) can be used to enhance 
the contrast of the cluster galaxies against the field galaxy 
population and, at the same time, to estimate the cluster redshift
(e.g. CIG J0848+4453  at z=1.273) \cite{sta97}. 

\noindent
\vskip 0.5cm

To summarize this first part of the review:
\vskip 0.5cm

Our knowledge of cluster properties (such as richness, morphology, 
space density, velocity dispersion, optical luminosity, morphological 
content etc.) and all correlations between these properties and their 
evolution (like Butcher-Oemler effect, space density as a function of 
redshift, optical morphology, etc.) has greatly improved in the last 
decade. The use of well-defined, automated  algorithms has allowed 
us to catalog a continuous range of galaxy overdensities from rich 
clusters down to poor groups. Near-infrared and radio selections
have pushed the detection of such systems to higher redshifts. The 
nearby surveys with their derived catalogs have contributed significantly 
to our understanding of the local universe and of the large-scale structure
and have allowed comparisons with galaxy clusters up  to large
look-back times ($z\geq1$). The modern 8m-10m telescopes are, and will be,
used to study the formation and evolution of the most distant galaxies and 
clusters.  By studying these objects, their mass, distribution and 
evolution we will obtain vital information on the formation history 
of the underlying mass field  which is a fundamental goal in cosmology. 

\section{Cluster Surveys and Catalogs in X-rays}
I have been advocating for many years the X-ray selection as 
an efficient method to find clusters of galaxies. The detection 
of X-ray emission is one of the cleanest ways to avoid sample 
contaminations, especially in the selection of high-z clusters. 
For the last several years X-ray cluster detection has been 
considered superior to classical optical detection methods because:

\noindent
$\bullet$ Clusters are powerful X-ray emitters with temperature 
and gas density which reflect the cluster gravitational potential

\noindent
$\bullet$  {\bf $L_{X}$} scales as the square of gas density so 
projection effects are minimized

\noindent
$\bullet$ Confusion from background fluctuations is much less prevalent
than in the optical

\noindent
$\bullet$  While in the optical the definition of cluster shape and mass 
with galaxy number counts and redshifts  is limited by the number of cluster 
galaxies, in X-rays the measurements are only limited by the number 
of photons received

\noindent
$\bullet$ Selection criteria are objective and quantifiable.

Naturally to unambiguously identify genuine distant clusters ($z>0.5$),
besides the evidence for hot gas through detection of X-ray emission, 
other criteria should hopefully be fullfilled like: measurement of projected 
overdensity of galaxies combined with significant overdensity observed in 
the redshift distribution, evidence for peak mass distribution, either
from weak or strong lensing and so on. Several new cluster lens 
systems have been discovered in X-ray selected samples \cite{lfev94}$^{,}$
\cite{lup99} and several weak lensing studies are using X-ray selected 
clusters (see \cite{mel99} for a review).

\begin{figure}
\begin{center}
\vskip +0.8truecm
{\psfig{figure=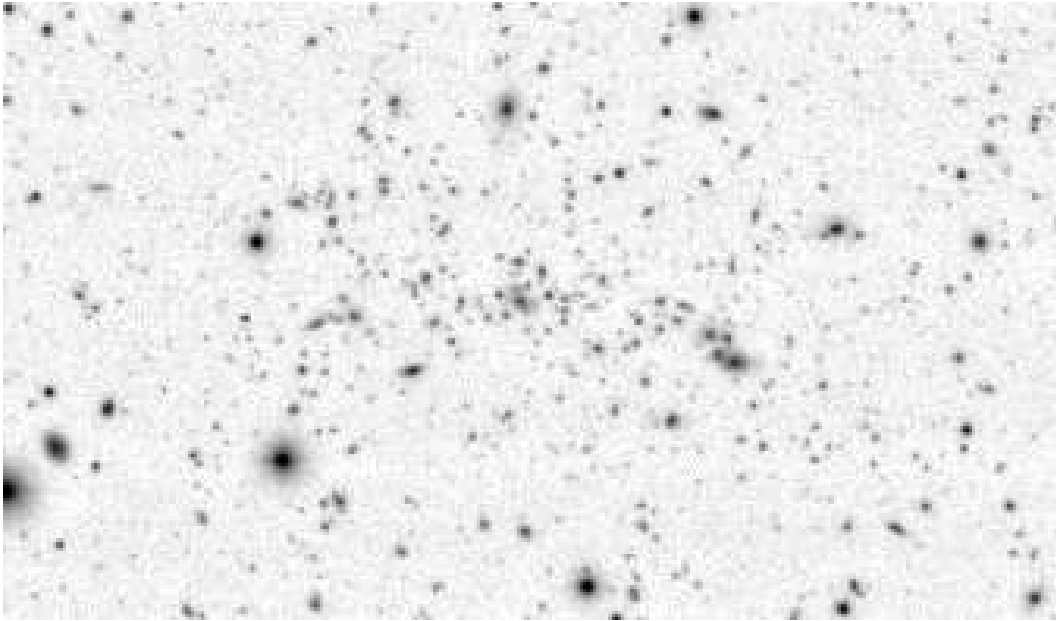, width=14 cm}}
\caption{Subarray I image of MS\thinspace 1054$-$03 at $z=0.83$
taken with the Tek 2048 CCD camera at the University of \hawaii 2.2m}
\end{center}
\end{figure}

The first X-ray imaging instruments onboard the Einstein Observatory
allowed the construction of the EMSS (Extended Medium Sensitivity 
Survey) catalog of clusters \cite{gio90a}$^{,}$ \cite{sto91}$^{,}$ 
\cite{hen92}. EMSS clusters are still observationally followed-up 
today. The cluster $MS\thinspace 1054-03$ \cite{gl94} is certainly one of the 
most amazing clusters known.  At a redshift of 0.83 \cite{gl94}, 
with a temperature of $12.4$ keV \cite{don98} and a mass of 
$7\times10^{14} h^{-1}$~M\sun \cite{don98}, the existence of massive 
clusters at $z\sim1$ is no longer in doubt (see Fig. 1).
All the X-ray surveys of the last decade, except for the EMSS,
have been carried out using ROSAT \cite{tru91} data, either from 
the All-Sky Survey (RASS \cite{vog99}) or from pointed observations. 
The surveys belonging to the former group, the contiguous area 
surveys, cover a very large solid angle ($\sim 10,000$\ \sq\deg\ or 
more, with the exception of the NEP \cite{gio00}) but are shallower 
than the pointed data surveys. The contiguous area
surveys  can be used to examine large-scale 
structure in the cluster distribution but they cannot detect in 
great number the most massive and luminous systems at high redshift 
($z>0.3$), with perhaps the exception of MACS \cite{eb00b} (see next section).
The great advantage of the serendipitous surveys, those extracted
from the pointed data, is their much deeper sensitivity, about two order 
of magnitude deeper than the contiguous area surveys
({$\sim 1\times10^{-14}$  vs $\sim 1\times10^{-12}$ erg cm$^{-2}$ s$^{-1}$
in the ROSAT band), even though their solid angle is less than 
$\sim200$\ \sq\deg, which makes them insensitive to massive clusters
in great number.

The REFLEX \cite{boe98}$^{,}$ \cite{guz99}, NORAS \cite{boe00}, 
BCS \cite{eb98} (and its extension eBCS \cite{eb00a}), 
RASS1-BS \cite{dg99}, MACS \cite{eb00b} and NEP \cite{bow96}$^{,}$ 
\cite{mul98}$^{,}$ \cite{gio00} belong to the group of 
contiguous area surveys. 
While the RDCS \cite{ros95}$^{,}$ \cite{ros98}, Southern SHARC 
\cite{bur97}$^{,}$ \cite{col97} and Bright SHARC \cite{rom00}, 
WARPS \cite{sch97}$^{,}$ \cite{jon98}$^{,}$ \cite{eb00c}, 
CFA 160\ \sq\deg\  \cite{vik98a}$^{,}$ \cite{vik00} and 
BMW \cite{laz99}$^{,}$ \cite{cam99} are serendipitous cluster surveys.

\section{The Contiguous Area Surveys}
The contiguous area surveys, and in particular the all sky surveys, 
are good tracers of large-scale structure given the large solid angle
sampled and the sizes of the superstructures (several hundreds
of Mpc). The selection through X-ray emission has the additional 
advantage of a more direct relation between luminosity and mass
and thus the derived X-ray luminosity function is most closely 
related to the mass function of the clusters which is used 
as an important calibrator of the amplitude of the density fluctuation 
power spectrum. The surveys described below, except for MACS,
sample the nearby universe ($z<0.3$) and are used as excellent 
reference for cluster studies at higher redshift. Deeper surveys, 
like the NEP, are not restricted to the local universe. The NEP 
has allowed the discovery of distant clusters, like 
$RX\thinspace J1716+66$ \cite{gio99}$^{,}$ \cite{hen97} at $z=0.81$,
a filamentary system very similar in many aspects
to the better known cluster  $MS\thinspace 1054-03$ (see Fig. 2).

\begin{figure}
\begin{center}
\vskip +0.8truecm
{\psfig{figure=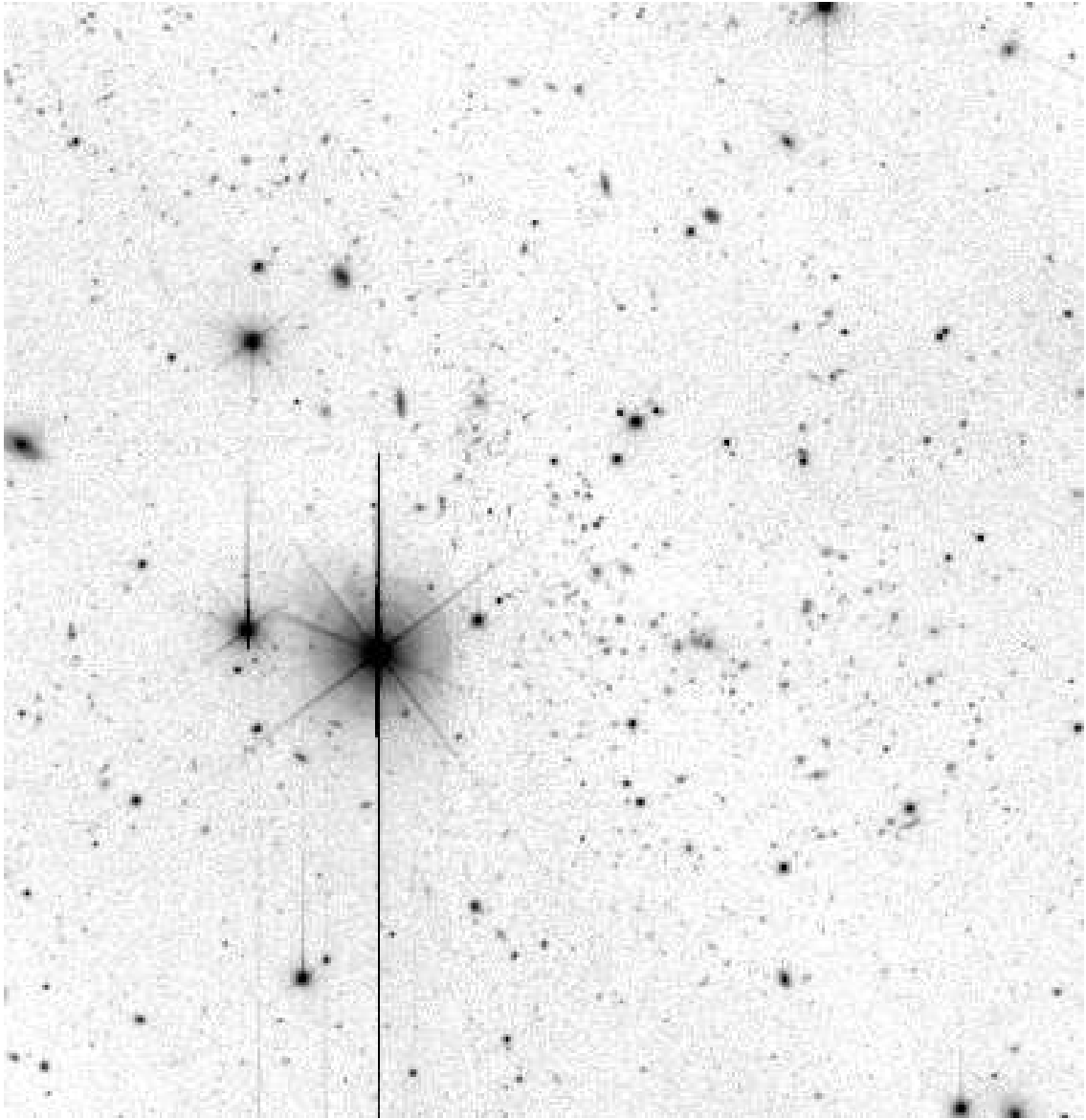, width=15 cm}}
\caption{Subarray I image of the NEP cluster $RXJ1716+66$ at
$z=0.81$ taken with the University of \hawaii $\times$8K CCD
mosaic camera on the CFHT prime focus}
\end{center}
\end{figure}

\subsection{The REFLEX and NORAS Galaxy Cluster Surveys}
There are two X-ray cluster surveys covering both celestial
hemispheres: the NOrthern Rosat All-Sky 
(NORAS \cite{boe00}) survey  and the complementary 
survey in the southern hemisphere, the REFLEX (Rosat Eso Flux 
Limited X-ray survey \cite{boe98}$^{,}$ \cite{guz99}). 
Both projects are carrying out the optical follow-up and 
redshift determination of clusters selected from the ROSAT 
All-Sky survey. While the NORAS 
cluster sample selection is purely based on X-ray information 
(e.g. extent of the X-ray source), the REFLEX survey is based on 
the correlation of X-ray sources with galaxy overdensities. 
Recent results from the REFLEX include a measure of the power
spectrum within a cubic volume of $400 h^{-1}$ Mpc side and 
an estimate of the two-point correlation function \cite{guz00}.
Hans B\"ohringer (these proceedings) will describe each survey 
in detail.

\subsection{The ROSAT Brightest Cluster sample (BCS), the 
RASS1 Bright Sample (RASS1-BS) and the Massive Cluster Survey (MACS)}}
\noindent The BCS, RASS1-BS and MACS cluster samples are all compiled from 
the ROSAT All-Sky Survey data. 

The BCS \cite{eb98} is an X-ray flux limited sample which
comprises 203 clusters in the northern hemisphere ($\delta \geq0$ 
and $|b| \geq 20\deg$) with measured redshifts $z\leq 0.3$. 
The sample is complete down 
to $4.4\times10^{-12}$ erg cm$^{-2}$ s$^{-1}$ in 0.1$-$2.4 keV.
The clusters have been selected from the RASS data using the 
Voronoi Tesselation and Percolation algorithm (VTP \cite{ew93})
optimized for the detection of irregular and extended X-ray 
sources.  The BCS  extension, the eBCS \cite{eb00a},  
comprises 107  X-ray selected clusters with fluxes 
$\geq 2.8\times10^{-12}$erg cm$^{-2}$ s$^{-1}$ (0.1$-$2.4 keV) and 
measured redshifts $z\leq 0.3$. Combining the two samples 
(310 clusters) there are signs of large-scale structure in the 
z-histogram at $z\sim0.036$ and $z\sim 0.077$ \cite{eb00a}.

The other survey of bright clusters  from the RASS is the RASS1 
Bright Sample \cite{dg99} in the southern hemisphere. 
The RASS1-BS is a flux limited sample of clusters that
uses the Steepness-Ratio Technique, SRT \cite{dg99}, which takes 
into account the extended nature of the X-ray emission of sources. 
The final  sample is count-rate 
limited and thus the completeness in flux varies between 
$\sim 3$ and $4\times10^{-12}$erg cm$^{-2}$ s$^{-1}$ in the hard  
$0.5-2$ keV band. It covers a total area of 
8324\ \sq\deg\ and includes 130 clusters of which  126  with 
measured z ($<0.3$). 
Both the BCS and the RASS1-BS have allowed an accurate determination of 
the bright end of the logN$-$logS and of the local cluster X-ray 
luminosity function over almost three decades in X-ray luminosity.
Both RASS samples provide an important reference 
for searches for cluster evolution at higher redshift.
There are of course many clusters in common between the last two
surveys and the previous section surveys since all of them are 
based on X-ray detections in the RASS.

The third cluster survey compiled from RASS data, 
the MACS \cite{eb00b}, is still  work in progress.  
The main characteristic of the survey is a very large solid angle 
($\sim23,000$\ \sq\deg) combined with a flux limit of 
$1\times10^{-12}$ erg cm$^{-2}$ s$^{-1}$ in  0.1$-$2.4 keV. Clusters
are selected from the RASS Bright Source Catalog 
(BSC \cite{vog99}) on the basis of the X-ray hardness ratio and with 
$|b| \geq 20\deg$ and $-40\deg \leq\delta\leq 80\deg$. The survey has 
provided so far the largest sample of clusters of galaxies at all 
redshifts (more than 850 clusters of which 755 with spectroscopic 
redshifts \cite{eb00b}). About 88 of such systems are in the 
range $0.3\leq z \leq0.6$ (see Fig. 3 for two MACS clusters at
$z=0.329$ and $z=0.57$).  Preliminary results 
from the MACS, based on a subsample of 25 X-ray brightest clusters,
show that negative evolution of the X-ray luminosity function 
is not significant at luminosities $L_X > 1\times 10^{45}$ erg s$^{-1}$ 
out to redshifts of $z\sim0.4$ \cite{eb00d} (see section 6 for 
a discussion on cluster evolution).

\noindent
\begin{figure}
\parbox{80mm}{\epsfig{file=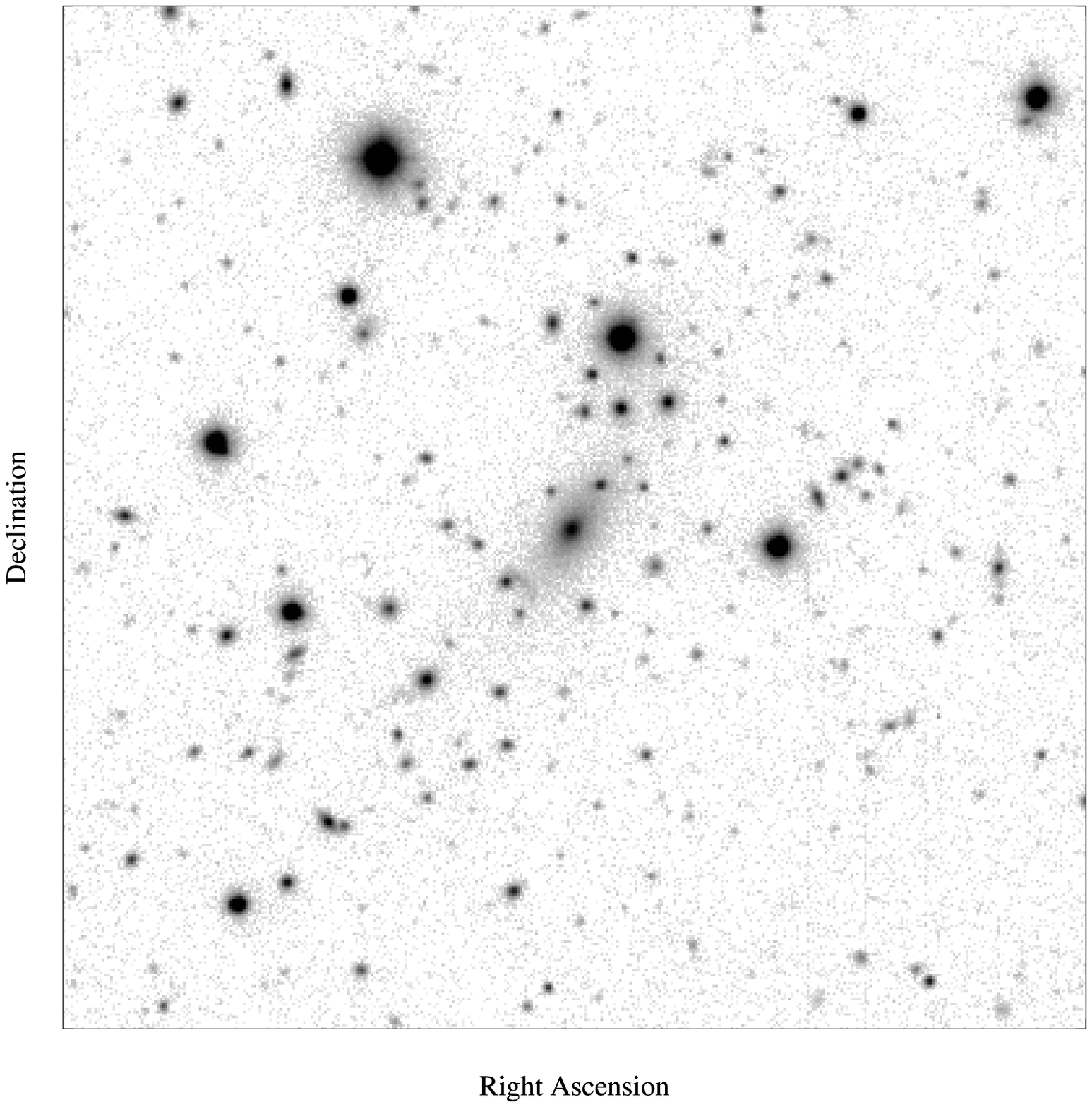,width=80mm}}
\parbox{80mm}{\epsfig{file=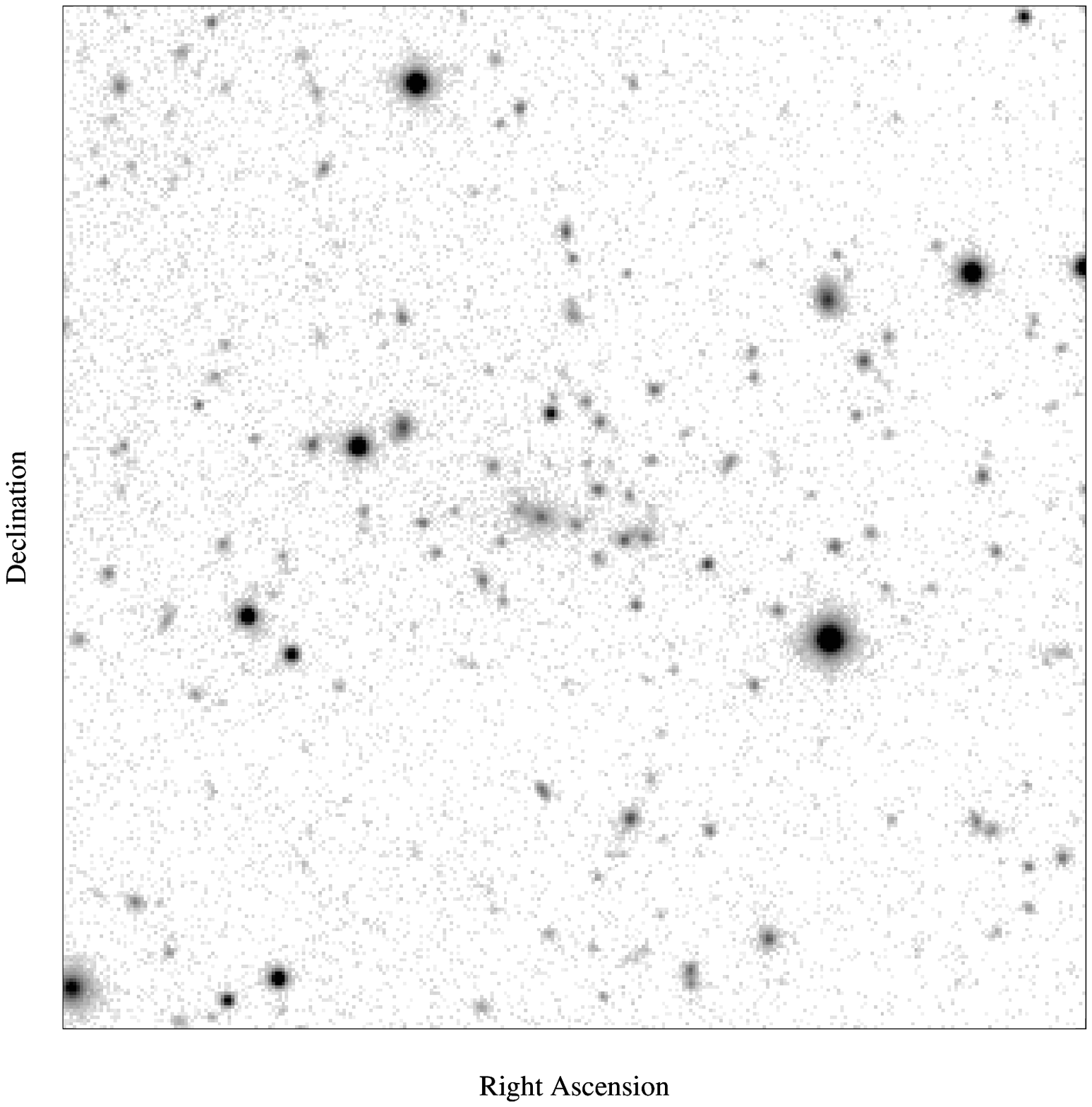,width=80mm}}
\caption{R band images of MACS clusters obtained with the University
of Hawaii's 2.2m telescope. Left: one of the most nearby and least
X-ray luminous clusters in the MACS sample ($z=0.329$, $L_{\rm X} =
6.4\times 10^{44}$ erg s$^{-1}$, 0.1--2.4 keV). Right: One of the most
distant and more X-ray luminous clusters in the MACS sample ($z=0.570$
$L_{\rm X} = 1.7\times 10^{45}$ erg s$^{-1}$, 0.1--2.4 keV). Both
clusters are new discoveries. The shown images span 1 $h_{\rm
50}^{-1}$ Mpc on the side at the cluster redshift.}
\end{figure}

\subsection{The North Ecliptic Pole survey (NEP)}
The last survey based on the ROSAT All$-$Sky survey data is the NEP 
\cite{hen94}$^{,}$ \cite{gio95}$^{,}$ \cite{mul98}$^{,}$ \cite{gio00},
a unique survey in its X-ray depth  and contiguous sky coverage. 
The NEP uses data from the deepest region of the RASS, 
where the scan circles converge and the effective exposure 
time approaches $40$ks, to produce a complete and unbiased 
X-ray selected sample of distant clusters. The survey covers
81\ \sq\deg\ down to a flux limit of $\sim3\times10^{-14}$ 
erg cm$^{-2}$ s$^{-1}$ in the $0.5-2.0$ keV band. 
There are 445 sources detected at 
$>4\sigma$ in the $0.1-2.4$ keV band. Sixty-four of these sources are 
identified with clusters, 19 of which have a redshift greater 
than 0.3 with the highest at z$=$0.81. The goal of the survey was
a better  understanding of the nature of cluster evolution
and a characterization of the three dimensional
large-scale structure of the universe by studying the
cluster-cluster correlation function.
A supercluster  of 21 groups and clusters has indeed 
been found during the analysis of the NEP sources \cite{mul00}.
Unlike most of the cluster surveys that are based  on 
X-ray cluster detections the NEP strategy is to identify 
through spectroscopic observations {\em all} the detected 
sources independently from their extension and nature. 
Comparing the number of the observed clusters in the NEP survey 
with the number of expected clusters assuming no$-$evolution  
models there is a deficit \cite{gio00}  of clusters with respect to
the local universe which is significant at $>4.7\sigma$. The
evolution appears to  commence at $L_{0.5-2}>1.8\times10^{44}$ erg
s$^{-1}$ in the NEP data. This result is in
agreement with the original EMSS results \cite{gio90b}$^{,}$ 
\cite{hen92} (see next section).

\section{The Serendipitous Cluster Surveys}
In the context of cluster evolution the X-ray luminosity function
(XLF) is a relevant quantity which parameterizes the number density 
of clusters per unit volume per luminosity interval.
The main result of the EMSS \cite{gio90a} cluster sample
was the evidence for cluster XLF evolution at the $3\sigma$ level. 
This evolution, dubbed negative evolution, goes in the sense that 
there are fewer \cite{gio90b}$^{,}$ \cite{hen92} distant clusters 
but only at  luminosities L$_{0.3-3.5} > 5\times10^{44}$ erg s$^{-1}$.
This behavior is expected in hierarchical formation theories where
the most massive clusters form last. This early result
inspired many EMSS$-$style cluster surveys,  
all based on ROSAT archival deep pointings. The issue 
of ``evolution versus no-evolution'' became a controversial and 
very hotly debated issue in the X-ray cluster community.
Each one of the  EMSS$-$style cluster surveys described below
covers an area of sky of less than 200\ \sq\deg, much less than the
$\sim800$\ \sq\deg\ of the original EMSS, but with  sensitivities
almost an order of magnitude deeper than the EMSS 
($\sim 1.8\times10^{-14}$ erg cm$^{-2}$ s$^{-1}$ vs
$\sim 1.3\times10^{-13}$ erg cm$^{-2}$ s$^{-1}$ in 0.3$-$3.5 keV).
Most of these surveys are still works in progress but a few
preliminary results are available. While everybody agrees
on the XLF of clusters with z$<$0.3 and 
$L_{(0.5-2)}<3\times10^{44}$ erg s$^{-1}$,
there is not a consensus yet for the XLF of the brightest  
and most distant  clusters known. As with the contiguous area 
surveys, several clusters are common to the surveys described 
below since all the surveys use the same pointed ROSAT data. 

\subsection{The Rosat Deep Cluster Survey (RDCS)} 
The RDCS \cite{ros95}$^{,}$ \cite{ros98} was designed to compile a large, 
X-ray flux-limited sample of galaxy clusters, selected via a serendipitous 
search for extended  X-ray sources in ROSAT-PSPC deep pointed observations.
The depth and the solid angle of the survey were chosen 
to probe an adequate range of X-ray luminosities over a large redshift 
baseline. Approximately 160 candidates were selected down to the flux 
limit of $1\times10^{-14}$ erg cm$^{-2}$ s$^{-1}$ in 0.5$-$2.0 keV, 
over a total area of 50\ \sq\deg, using a wavelet detection algorithm.
This technique is particularly efficient in discriminating between
pointlike and extended, low-surface-brightness sources \cite{ros95}. 
The RDCS has not been completely identified yet but complete subsamples 
have been created to allow a statistical reconnaissance of the data. 
The most interesting finding, beside the detection of an 
X-ray selected cluster at z=1.26 ($RX\thinspace J0848.9+4452$ \cite{ros99}), 
is the confirmation of the EMSS results: the
most luminous, presumably most massive clusters are indeed rarer at 
high redshift \cite{ros00}$^{,}$ \cite{bor00}.

\subsection{The Wide Angle Pointed Rosat Survey (WARPS)}
The goal of the WARPS, as with all the other serendipitous surveys,
was to compile a complete, unbiased, X-ray selected
sample of clusters of galaxies from serendipitous detections
in deep, high-latitude ROSAT PSPC pointings \cite{sch97}$^{,}$ 
\cite{jon98}. The detection algorithm employed is
the VTP algorithm, particularly suited for detection of
low-surface-brightness emission sources. Optical follow-up is
not restricted to the obvious extended sources but includes
also point-like sources in order to be as complete
as possible in the detection of possible unresolved clusters.
The survey is still work in progress and as of today the
WARPS covers 73\ \sq\deg\ down to a flux limit of
$5.5\times10^{-14}$  erg cm$^{-2}$ s$^{-1}$ in 
0.5$-$2.0 keV \cite{eb00c}. There are about 150 clusters, 
half of which at $z>0.3$, in the luminosity range from 
$10^{42}$ to $8\times10^{44}$ erg s$^{-1}$.
A comparison of the WARPS XLF with the local BCS luminosity function
does not show any evidence for luminosity evolution up to 
z=0.83 \cite{jon00}.

\subsection{The Serendipitous High-Redshift Archival Rosat Cluster 
surveys (Southern SHARC and Bright SHARC)}
The southern SHARC survey consists of 35 clusters of galaxies
discovered as extended sources at a flux limit of 
$\simeq 3.9\times10^{-14}$ erg cm$^{-2}$ s$^{-1}$ in 0.5$-$2 keV
in less than 20\ \sq\deg\ of sky \cite{col97}. There are 16 clusters 
in the redshift range $0.3<z<0.7$ detected in 66 deep ROSAT PSPC pointings
\cite{bur97}. No evidence for evolution \cite{col97}$^{,}$ 
\cite{bur97} of the XLF  over the luminosity range
$L_{X}\sim(0.3-3)\times10^{44}$ erg s$^{-1}$ in 0.5$-$2.0 keV
is found, in  agreement with the XLF of the low redshift cluster 
population. However, the SHARC sample is not able to
test the high luminosity-high redshift space where the XLF 
evolution was detected.

The Bright SHARC survey \cite{rom00} is the largest 
of the serendipitous surveys covering 178\ \sq\deg\ of sky. 
The survey uses a wavelet based algorithm
and detects 374 extended sources in deep (t$_{exp}>10$ks) PSPC pointings. 
The brightest 94 sources above a flux limit of
$1.4\times10^{-13}$ erg cm$^{-2}$ s$^{-1}$  in 0.5$-$2 keV
constitute the Bright SHARC cluster candidate sample.
Optical follow-up has identified 37 clusters of which 12 in the 
range $0.3<z<0.83$. Adding to the SHARC sample
the 160\ \sq\deg\ sample (see below) the authors \cite{nic99} 
present evidence for a deficit of clusters at $L_{X} > 5\times10^{44}$
erg s$^{-1}$  compared to what is expected from a non-evolving XLF.

\subsection{The CFA 160\ \sq\deg\ survey}
The 160\ \sq \deg ~X-ray catalog \cite{vik98a} was constructed 
by detecting  significantly extended X-ray sources in 653 ROSAT 
PSPC pointed observations. After the Bright SHARC it is among the 
largest of the new serendipitous surveys 
and comprises 203 clusters (with redshift measurements for 100 
X-ray brightest clusters \cite{vik00}). The X-ray detection is 
performed in a fully automated and objective way and 
selection effects are well understood. Given the large area combined
with high sensitivity the 160\ \sq\deg\ survey is particularly suited 
to probe the evolution of the bright end of the cluster 
XLF ($L_{X}>3\times10^{44}$ erg s$^{-1}$ in $0.5-2$ keV band).
A calculation of the XLF in the redshift interval $0.3<z<0.8$
(37 clusters with $f_{X}>1.4\times10^{-14}$ and  
$L_{X} > 0.71\times10^{44}$ erg s$^{-1}$) shows a deficit \cite{vik00} 
of high luminosity clusters in agreement with the
EMSS XLF \cite{gio90b}$^{,}$ \cite{hen92}.

\subsection{The Brera Multiscale Wavelet survey (BMW)} 
The BMW \cite{laz99}$^{,}$ \cite{cam99} is the most recently 
initiated survey  and it is still work in
progress. The BMW is the only serendipitous survey which 
applies a wavelet detection  algorithm, first used
by the RDCS, to analyze images taken with 
the ROSAT HRI instrument rather than the PSPC. 
Analysis of a large set of HRI data will allow 
the investigators to survey $\sim400$\ \sq\deg\ down to a limiting
flux of $\sim 1\times10^{-13}$ erg s$^{-1}$ cm$^{-2}$, and a much
smaller volume ($\sim0.3$\ \sq\deg) down to 
$\sim 3\times10^{-15}$ erg cm$^{-2}$ s$^{-1}$. A complete catalog will
result from their analysis, consisting of the BMW 
Bright Source Catalog (BMW-BSC), with
sources detected with a significance of $\geq4.5 \sigma$ and the Faint 
Source Catalog (BMW-FSC), with sources at $\geq3.5 \sigma$. 
At present there are about 300 cluster candidates (Guzzo, private 
communication) which are being imaged at optical telescopes for
confirmation. The survey is at a too early stage to give any
indication of the behavior of the high-luminosity, high-redshift 
galaxy clusters.

\begin{figure}
\begin{center}
\vskip +0.8truecm
{\psfig{figure=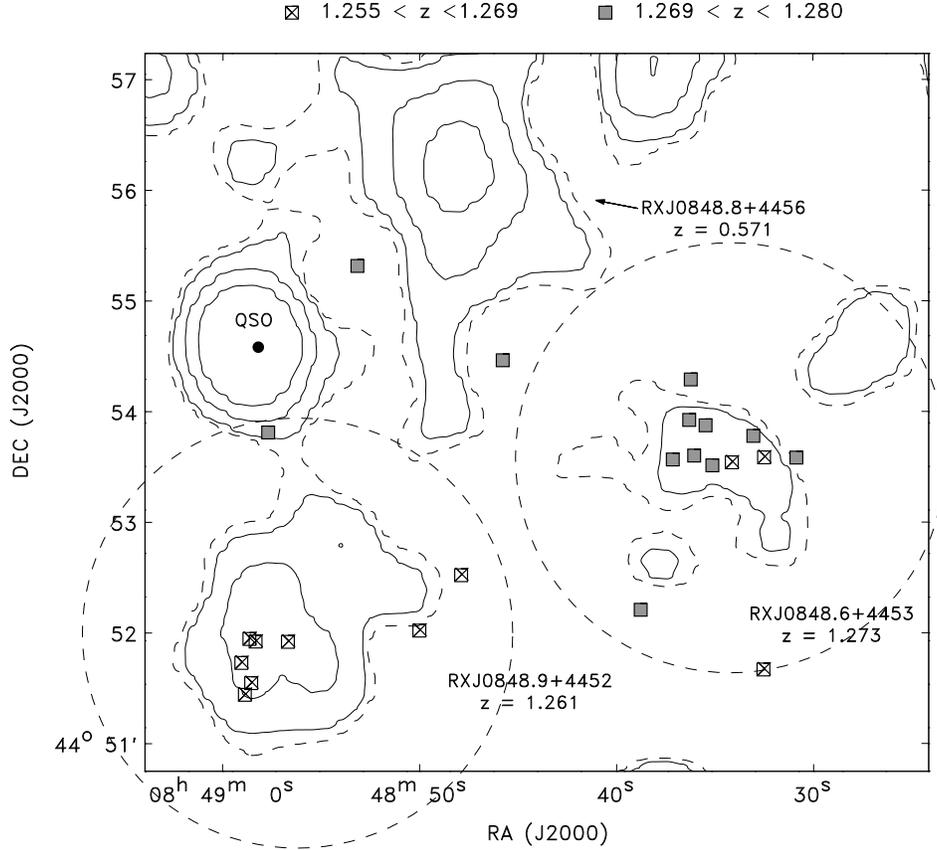, width=13 cm}}
\caption{Adapted from Fig. 6 in Rosati \etal 1999; 
original caption reads: Lynx field overlaid ROSAT-PSPC contours and 
spectroscopic members. The circles are centered at the X-ray centroids 
of the two clusters with a radius of $116\arcmin=1 h^{-1}_{50} Mpc$ 
at $z=1.27$. Filled squares refer to clusters with $1.255<z<1.269$,
and squares with crosses refer to clusters with $1.269<z<1.280$.}
\end{center}
\end{figure}

\section{Final Remarks}
Many statistical and individual cluster studies have 
been made in the last decade. This has been possible thanks
to the operation of larger and more sensitive telescopes either  
ground based and in space. 
In X-rays more accurate observable and derived cluster properties (like
flux, luminosity, gas density profiles, gas temperatures, cluster 
gas and total mass, etc.) and their relations (like
the luminosity-temperature or the mass-temperature relations)
are now available.  Mass determinations from studies of velocity 
dispersion and from weak lensing are in good agreement with 
mass derivations from X-ray temparatures measurements.
We have a better understanding of the effect
of the environment on cluster properties (i.e. presence of cooling 
flows, extended radio lobes, clumpiness seen in optical and IR, 
morphological content of galaxies, etc.). 

The sensitivity of the X-ray surveys is comparable to other
wavelength sensitivities and clusters can be detected out to 
$z>1$ \cite{ros99}. Detailed studies
of high-z massive clusters have shown that they are filamentary in optical
with the X-ray emission  following the elongation of the optical galaxies
\cite{don98}$^{,}$ \cite{gio99}$^{,}$ \cite{eb00c}. Most high-z clusters
have high velocity dispersions (when many redshift measurements are 
available)  as well as high gas temperatures \cite{don98}$^{,}$ \cite{gio00a}.
Three out of four X-ray selected clusters at $z\sim0.8$ are not relaxed
($RX\thinspace J1716+66$ \cite{gio99}$^{,}$ \cite{clo98}$^{,}$
\cite{hen97} at $z=0.81$;
$MS\thinspace 1054-03$ \cite{don98}$^{,}$ \cite{lk97} at $z=0.83$; 
$RX\thinspace J0152-13$ \cite{dec00}$^{,}$ \cite{eb00c} at $z=0.83$ and 
$MS\thinspace 1137+66$ \cite{don99}$^{,}$ \cite{clo00} 
at $z=0.78$, this last one is well compact and unrelaxed). 
The rarity of relaxed systems in high-z clusters may indicate 
that we are beginning to observe the formation epoch of the majority
of massive clusters. 

Progress has been made in the detection of large-scale
structure both nearby \cite{mul00} and out to $z\sim1.27$ 
(see Fig. 4 here and the beautiful color image of the Lynx 
field in Rosati et al. 1999 \cite{ros99}, their Fig. 7). 
Also a possible detection  of X-ray emission from an intra-cluster 
filament has been reported \cite{sch00}. 

There is accumulating evidence in favor of negative evolution, that is 
less high-z high luminous clusters in the past. All the new
serendipitous surveys, except for the WARPS, are now reporting mild negative
evolution at varying levels of significance. However, given the small
number statistics and incomplete optical follow-ups of some existing
serendipitous clusters surveys, the issue has not been completely
resolved yet.  The  high throughput and energy resolution of the
large X-ray telescopes now in orbit,
Chandra and XMM-Newton, will allow to obtain 
more reliable flux measurements which will help reduce some of the
systematic uncertainties in the derived XLF and number counts of
clusters of galaxies.

\section*{Acknowledgments}
Partial financial support from  NSF grants AST91-19216 and
AST-95-00515, and  from CNR-ASI grants are gratefully 
acknowledgd. I wish to thank Florence, Daniel and 
Mme Raban for organizing a great meeting in always
charming Paris.

\end{document}